\documentclass[conference]{IEEEtran}
\IEEEoverridecommandlockouts
\usepackage{cite}
\usepackage{amsmath,amssymb,amsfonts}
\usepackage{algorithmic}
\usepackage{graphicx}
\usepackage{textcomp}
\usepackage{xcolor}
\usepackage[a4paper, total={184mm,239mm}]{geometry}
\newcommand{\eat}[1]{}
\usepackage{amsthm}
\usepackage{cleveref}
\usepackage{subfigure}
\usepackage[ruled,linesnumbered]{algorithm2e}
\usepackage{multirow}
\crefname{figure}{Fig.}{Fig.}
\crefname{table}{Table}{Table}
\def\BibTeX{{\rm B\kern-.05em{\sc i\kern-.025em b}\kern-.08em
    T\kern-.1667em\lower.7ex\hbox{E}\kern-.125emX}}

\begin{document}

% \title{Conference Paper Title*\\
% {\footnotesize \textsuperscript{*}Note: Sub-titles are not captured in Xplore and
% should not be used}
% \thanks{Identify applicable funding agency here. If none, delete this.}
% }
\title{GPU-Accelerated Effective Resistance Analysis for 3D IC Power Delivery Network}

% \author{\IEEEauthorblockN{1\textsuperscript{st} Given Name Surname}
% \IEEEauthorblockA{\textit{dept. name of organization (of Aff.)} \\
% \textit{name of organization (of Aff.)}\\
% City, Country \\
% email address or ORCID}
% \and
% \IEEEauthorblockN{2\textsuperscript{nd} Given Name Surname}
% \IEEEauthorblockA{\textit{dept. name of organization (of Aff.)} \\
% \textit{name of organization (of Aff.)}\\
% City, Country \\
% email address or ORCID}
% \and
% \IEEEauthorblockN{3\textsuperscript{rd} Given Name Surname}
% \IEEEauthorblockA{\textit{dept. name of organization (of Aff.)} \\
% \textit{name of organization (of Aff.)}\\
% City, Country \\
% email address or ORCID}
% \and
% \IEEEauthorblockN{4\textsuperscript{th} Given Name Surname}
% \IEEEauthorblockA{\textit{dept. name of organization (of Aff.)} \\
% \textit{name of organization (of Aff.)}\\
% City, Country \\
% email address or ORCID}
% \and
% \IEEEauthorblockN{5\textsuperscript{th} Given Name Surname}
% \IEEEauthorblockA{\textit{dept. name of organization (of Aff.)} \\
% \textit{name of organization (of Aff.)}\\
% City, Country \\
% email address or ORCID}
% \and
% \IEEEauthorblockN{6\textsuperscript{th} Given Name Surname}
% \IEEEauthorblockA{\textit{dept. name of organization (of Aff.)} \\
% \textit{name of organization (of Aff.)}\\
% City, Country \\
% email address or ORCID}
% }

\author{
    \IEEEauthorblockN{Jingchao Hu, Cheng Zhuo, Zhou Jin}
    \IEEEauthorblockA{
        College of Integrated Circuit, Zhejiang University
    }
}

\maketitle

\begin{abstract}
    Three-dimensional (3D) integration is a critical technique for enhancing transistor density, improving power efficiency, and reducing interconnect delays. However, as current demands and design complexity increase, power deliver networks (PDNs) are facing growing challenges.
    Careful planning of through-silicon vias (TSVs) is essential for ensuring reliable PDNs, where effective resistance serves as a vital metric for the reliability. Ill-planned TSVs often cause 3D IC with unevenly distributed effective resistance and consequently severer IR Drop.
    In this paper, we propose a GPU-accelerated framework on accurate effective resistance analysis for early stage 3D IC PDNs. The proposed framework achieves a speedup of 5 to 6 orders of magnitude compared to the conventional direct solver, while maintaining negligible deviations in both maximum and average relative errors.\vspace{-0.2cm}

\end{abstract}

%\begin{IEEEkeywords}
%Early-stage, effective resistance, PDN, GPU
%\end{IEEEkeywords}

% \section{Introduction}
% This document is a model and instructions for \LaTeX.
% Please observe the conference page limits.
\section{Introduction}
\label{Sec:Introduction}

As the semiconductor industry continues to push the boundaries of Moore's Law, three-dimensional (3D) integration has emerged as a vital technology for increasing transistor density in integrated circuits (ICs). By utilizing through-silicon vias (TSVs) to vertically stack multiple dies, 3D integration provides several key advantages, such as increased bandwidth, reduced footprint, lower interconnect delay, and improved energy efficiency~\cite{3D_Intel}. However, the implementation of 3D ICs introduces substantial challenges, particularly in the design and verification of power delivery networks (PDNs)\cite{Design_of_reliable_3D_IC, 2D-3D-PDN-comparison}. Unlike traditional two-dimensional (2D) ICs, where each die has an independent PDN, 3D ICs employ a shared PDN that connects multiple dies through TSVs. This shared PDN architecture significantly increases design complexity, exacerbating issues such as power supply noise (PSN) and thermal hotspots. As a result, the risk of excessive IR drop becomes a critical concern, complicating the reliable design of 3D IC PDNs\cite{Analysis_of_IR_drop}. %With narrower noise margins, excessive IR drop in the PDN becomes a critical issue, potentially causing the supply voltage to dip below the threshold, leading to functional errors or even chip failure~\cite{Analysis_of_IR_drop}.

%In 3D ICs, the floorplanning and grouping of TSVs have a significant impact on the PDN's ability to maintain stable voltage levels across all dies~\cite{TSV-Based-3-D-ICs}. TSV optimization and placement is often determined early in the design process due to its implications for both packaging and manufacturing. Once fixed, these TSVs define the current paths within the PDN, making it crucial to ensure that their placement supports the reliability and performance of 3D ICs. \textit{Effective resistance} serves as an important metric for assessing how current flows through the PDN and how voltage drops are distributed across the network~\cite{XGBIR, IncPIRD}. Uneven resistance distribution can exacerbate IR drop issues, particularly in regions with high current density or inadequate connectivity, which cannot be easily resolved at latter design stages. Thus, \textbf{the capability of efficiently analyzing and optimizing effective resistance at early design stage is highly desired for 3D IC designers.}

In 3D ICs, maintaining stable voltage levels across all dies is critical, and this stability is heavily influenced by the placement and grouping of TSVs~\cite{TSV-Based-3-D-ICs}. TSV placement is often determined early in the design process due to its implications for both packaging and manufacturing. Once fixed, these TSVs define the current paths within the PDN, making it crucial to ensure that their placement supports the reliability and performance of 3D ICs. \textit{Effective resistance} plays a key role in evaluating current flow through the PDN and the distribution of voltage drops~\cite{XGBIR, IncPIRD}. Uneven resistance distribution can lead to localized IR drops that degrade performance or cause chip failure, which are difficult to resolve in later design stages. Thus, \textbf{fast and accurate effective resistance analysis is crucial during the early stages of 3D IC design}.

Effective resistance analysis has traditionally been approached as part of power integrity verification or PSN sign-off using standard PSN analysis methods~\cite{random_walk, Kozhaya_Nassif_Najm_2001, Wu_2004, hierarchical_analysis,  effR-approx-infinity-mirror-technique, effR_of_2layers, effR_approach_for_PG, effR-for-PG-TSI, GPU_based_PDN_analysis, chen2008algorithm, effR-sparse-cholesky-inverse}. However, \textit{most existing methods were developed for 2D PDNs and do not fully leverage the structural properties of 3D PDNs, potentially resulting in inefficient computations when applied to 3D ICs.}
Moreover, early-stage PDN design requires the ability to rapidly evaluate various design options~\cite{early_stage_analysis}. As the number of stacked dies increases, the number of possible effective resistance paths from the IPs/modules on the bottom tier to the bumps on the top tier also increases exponentially, making traditional methods for analyzing effective resistance computationally expensive and slow. At the early design stage, there exist tens to hundreds of thousands of effective resistance paths from bumps, through TSVs, to the current loads (defined at IPs or modules) in 3D ICs. Although some approaches have explored parallel acceleration techniques for general PDN analysis, these methods are \textit{not optimized for effective resistance calculations, particularly for the unique structure of 3D ICs}.

To address these challenges, this paper introduces an advanced GPU-accelerated framework tailored for early-stage effective resistance analysis in 3D IC PDNs. The proposed framework enhances the computation of effective resistance across various bump-load configurations by employing a divide-and-conquer strategy that partitions the 3D PDN into multiple 2D sub-networks. This decomposition not only enables efficient GPU-accelerated analysis but also significantly reduces computational complexity, thereby enhancing scalability for larger 3D IC designs. The contributions are summarized as follows:
%Thus, to address these issues, this paper proposes an framework for early-stage analysis and optimization of TSV planning in 3D ICs. This tool enables designers to achieve reliable planning early in the design process, thereby avoiding costly optimizations later. The main contributions of this paper include:
\begin{itemize}
 \item We \textbf{introduce a divide-and-conquer strategy} that decomposes the 3D PDN into multiple 2D sub-problems, which can be analyzed independently and in parallel to improve the scalability.
    \item We \textbf{reformulate the original PDN system conductance matrix}—initially intended for voltage drop analysis—into a specialized formulation optimized for effective resistance computation, thereby reducing the problem size.
    
    % Even at the early stage, where current loads are defined at the IP or module level, there are still tens to hundreds of thousands of effective resistance paths from bumps through TSVs to current loads in 3D ICs. To efficiently compute these effective resistances, we reformulate the system conductance matrix from MNA and reduce the linear solve to a much smaller problem than typical PDN analysis, which is key to achieving speedup. 
    %Even at the early stage, where current loads are defined at the IP or module level, there are still tens to hundreds of thousands of effective resistance paths from bumps through TSVs to current loads in 3D ICs. To efficiently compute these effective resistances, we reformulate the system conductance matrix and reduce the problem size, enabling faster computation compared to traditional methods.
   
    %We propose a multigrid-based divide-and-conquer strategy to efficiently map a 3D PDN analysis problem to multiple 2D cases. Combined with the restructured formulation, this approach allows parallel computation, enhancing scalability for larger 3D IC designs.
    %We propose a divide-and-conquer strategy that decomposes the 3D PDN into multiple 2D sub-problems, which can be analyzed independently and in parallel, enhancing scalability and reducing computation time.
    \item  The proposed framework \textbf{is highly compatible with GPU-acceleration platforms}, enabling concurrent computation of effective resistance across multiple bump-load pairs, thus further accelerating the analysis process.
    %Using the given optimization targets and available design parameters, we employ a Bayesian optimization method to identify the optimal combination of possible TSV physical parameters and allocation, significantly reducing designers' turnaround time.
    %Our approach is well-suited for parallel computing platforms, enabling simultaneous computation of effective resistance across multiple bump-load combinations, further accelerating the analysis. 
\end{itemize}
The experimental results demonstrate that the proposed framework significantly outperforms the prior works in both efficiency and scalability. For large-scale 3D IC designs, our framework was able to compute effective resistances for millions of bump-load pairs in a matter of seconds, a task that would be nearly infeasible using a golden direct solver~\cite{chen2008algorithm} or a most recently published effective resistance solver~\cite{effR-sparse-cholesky-inverse}. The results also highlight the scalability of the proposed framework, with runtimes that increase linearly with the total number of nodes and bump-load pairs, making it well-suited for modern 3D PDN designs. Finally, our framework allows designers to explore and optimize TSV planning, facilitating better current distribution or significant routing area saving in practical scenarios. %These advantages position the proposed framework as an efficient and practical tool for early-stage PDN analysis and optimization in 3D ICs.
\section{Background}
\label{Sec:Background}
% \subsection{Geometric Multi-grid}
\subsection{Formulation of DC analysis for PDN}

% \begin{figure}[htbp]
%     \centering
%     \includegraphics[width=0.5\linewidth]{figs/equivalent_circuit.png}
%     \caption{Equivalent circuit of 3D PDN (rough version)}
%     \label{fig:eq_circuit}
% \end{figure}
% For DC analysis, 3D PDN can be modeled as the equivalent circuit as illustrated in \cref{fig:eq_circuit} and according to modified nodal analysis (MNA), the DC analysis can be formulated as:
Power grid analysis is essential for evaluating voltage drops across nodes in the PDN. In DC analysis, the power grid is modeled as a resistive network, represented by the following linear system using modified nodal analysis (MNA):
\begin{equation}\vspace{-1ex}
    GV = I\label{eq:DC_formulation}
\end{equation}
where $G$ is the conductance matrix, $V$ is the vector of node voltages, and $I$ is the vector of current sources. Given that $G$ is sparse, symmetric, and positive definite, direct solution methods typically utilize Cholesky factorization with forward/backward substitution to solve this system~\cite{direct_solver_for_pg, chen2008algorithm}. DC analysis is routinely performed to compute the voltages of all PDN nodes, from early-stage checks to sign-off verification~\cite{chen2008algorithm, early_stage_analysis}.

\subsection{Effective resistance of PDN}

\begin{figure}[t]
    \centering\vspace{-0.5cm}
    \includegraphics[width=0.5\linewidth]{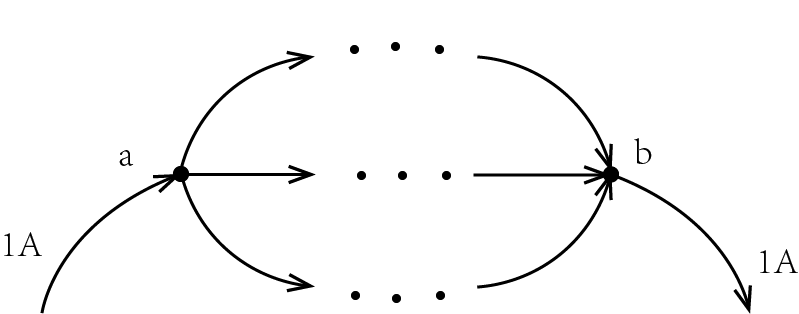}\vspace{-0.3cm}
    \caption{Definition of effective resistance~\cite{effR_definition}.}
    \label{fig:effr-definition}\vspace{-0.3cm}
\end{figure}

In a PDN, the effective resistance between two nodes, $a$ and $b$, can be defined as the absolute value of the voltage difference between them when a unit current is injected into $a$ and extracted from $b$, as shown in \cref{fig:effr-definition}~\cite{effR_definition}. Let $G$ be a graph representing the PDN with its Laplacian matrix $L_G$. The effective resistance between $a$ and $b$ is: 
\begin{equation} \label{eq.the general approch of effr computation} R(a,b) = (e_a - e_b)^T L_G^\dagger (e_a - e_b) \end{equation} 
where $L_G^\dag$ is the pseudo-inverse of $L_G$, and $e_a$ denotes the $a_{th}$ column of the identity matrix. As this expression suggests, computing effective resistance in general is computationally expensive due to the matrix inversion involved. 

For early stage, it is critical to find out the effective resistance from bumps to current loads for TSV or bump planning. There are a few prior works focusing on simplifying the above formulation from the definition. For example, for an early-stage 2D uniform PDN with $M\times N$ nodes, \cite{Wu_2004} proposed to calculate the effective resistance across $(x_1, y_1)$ and $(x_2, y_2)$ as:
%\begin{small}
\begin{equation}
\small
\label{eq:effR_for_regular_grid}
\begin{aligned}
&R = \frac{r}{N}|x_1-x_2|+\frac{s}{M}|y_1-y_2|+ 
        \frac{2}{MN}\sum_{m=1}^{M-1}\sum_{n=1}^{N-1} \\
        &\frac{
            [
                cos(x_1 + \frac{1}{2})\theta_m cos(y_1 + \frac{1}{2})\phi_n - 
                cos(x_2 + \frac{1}{2})\theta_m cos(y_2 + \frac{1}{2})\phi_n
            ]^2
        }
        {r^{-1}(1-cos\theta_m)+s^{-1}(1-cos\phi_n)}
\end{aligned}  
\end{equation}
\hspace{-1px}where $r$ and $s$ are the unit resistance for the horizontally- and vertically routed PDN wires, 
$\theta_m=\frac{m\pi}{M}$ and ${\phi_n=\frac{n\pi}{N}}$.

With the computed effective resistance, designers can easily compute the voltage at critical nodes, which is more important at early design stage than obtaining the voltages at all nodes~\cite{early_stage_analysis, effR_approach_for_PG}. For instance, as shown in \cite{effR_approach_for_PG}, if the bump node $n_{{bump}}$ is connected to an ideal voltage $v_{{bump}}$ and $n_{load}$ nodes are connected to current loads with the $k_{th}$ load carrying $I_{{load}(k)}$, the node voltage $v_{j}$ at node $n_j$ is expressed by:
\begin{equation}\begin{aligned}
\label{eq:effR based node voltage}
    v_{j}=v_{bump}-\frac{1}{2}\sum_{k=1}^{n_{load}}
    [I_{load(k)}\times(R_{n_{bump}, n_j}\\
    +R_{n_{bump}, n_{load(k)}}-R_{n_j, n_{load(k)}})]
\end{aligned}\end{equation}
where $R_{n_{bump}, n_j}$, $R_{n_{bump}, n_{load(k)}}$, and $R_{n_j, n_{load(k)}}$ are the effective resistances between $n_{bump}$ and $n_j$, $n_{bump}$ and $n_{load(k)}$, and $n_j$ and $n_{load(k)}$, respectively.

\section{Proposed Framework}
\label{Sec:Framework}
\subsection{Overview}
% \begin{figure}[htbp]
% \centerline{\includegraphics{figs/flow.png}}
% \caption{Example of a figure caption.}
% \label{fig}
% \end{figure}
\begin{figure*}[htbp]\vspace{-0.5cm}
    \centering
    \subfigure[Synthesized layout of 3D IC PDN]{
        \label{fig:layout}
        \includegraphics[width=0.22\linewidth]{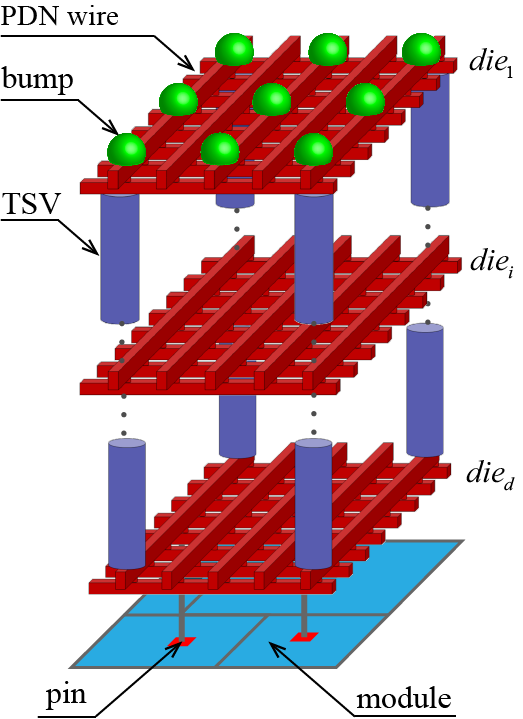}
    }
    \subfigure[Equivalent circuit model extraction]{
        \label{fig:netlist}
        \includegraphics[width=0.22\linewidth]{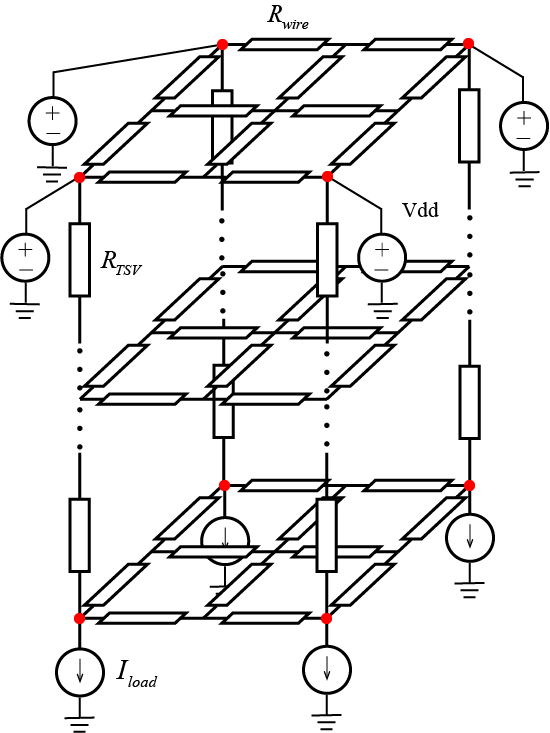}
    }
    % \subfigure[die-stacked structure (rough version)]{
    %     \label{fig:structure}
    %     \includegraphics[width=0.6\linewidth]{figs/structure.png}
    % }
    % \subfigure[introduction of unit current]{
    %     \label{fig:unit_current}
    %     \includegraphics[width=0.45\linewidth]{figs/unit_current.png}
    % }
    \subfigure[Divide-and-conquer strategy]{
        \label{fig:n-die PDN with unit current}
        \includegraphics[width=0.22\linewidth]{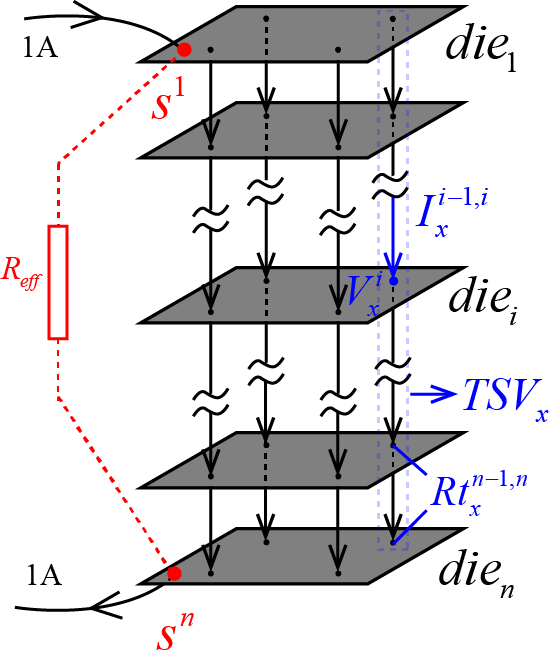}
    }
    % \subfigure[$die_1$]{
    %     \label{fig:die1}
    %     \includegraphics[width=0.35\linewidth]{figs/die-by-die1.png}
    % }
    % \subfigure[Inter-die effective resistance]{
    \subfigure[TSV current merging]{
        \label{fig:tsv-model}
        \includegraphics[width=0.22\linewidth]{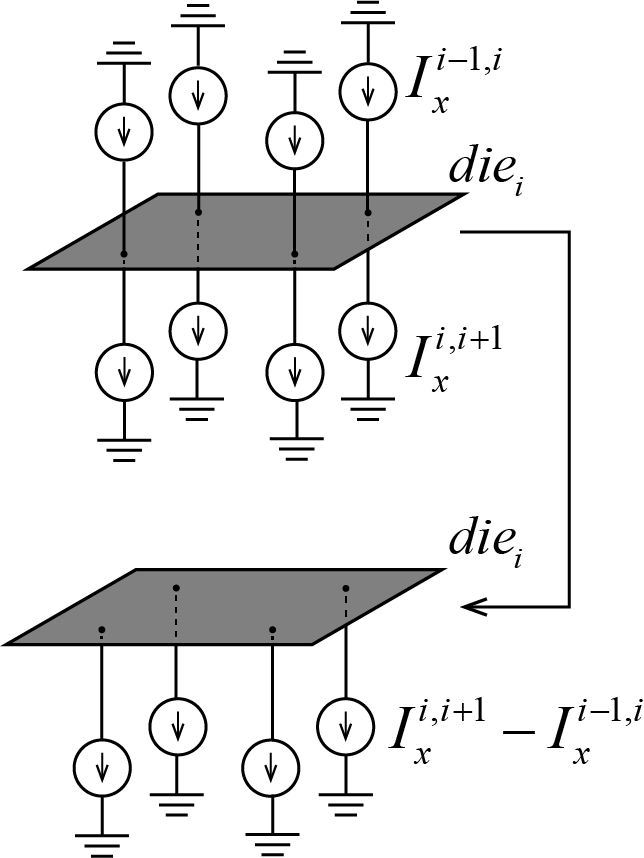}
    }
    % \subfigure[$die_2$]{
    %     \label{fig:die2}
    %     \includegraphics[width=0.55\linewidth]{figs/die-by-die2.png}
    % }
    % \includegraphics[width=0.5\linewidth]{}
    \vspace{-0.3cm}\caption{Key steps of the proposed framework for effective resistance analysis of 3D PDN.}\vspace{-0.5cm}
    \label{fig:enter-label}
\end{figure*}

%Effective resistance analysis is a critical task during the early stages of 3D IC PDN design~\cite{}. At this stage, designers need to quickly evaluate the effective resistance from current loads to bumps for various TSV or bump planning scenarios. 

In the conventional design flow, PDN is typically finalized during the placement and routing stage.  Thus, at the early design stage, designers synthesize the PDN using a few critical physical parameters, such as wire width and pitch for each layer, TSV size, pitch, and density, bump pitch, and density, $etc.$, to generate a synthetic layout, as shown in Fig.~\ref{fig:layout}. This layout is then extracted to build the equivalent resistor network model of the PDN (Fig.~\ref{fig:netlist}). Although early-stage PDNs often exhibit uniform mesh structures that can be exploited to expedite analysis, the sheer size and complexity of 3D ICs make even these simplified PDNs computationally challenging to analyze. To overcome these challenges, the proposed framework computes the effective resistance through the following:
\begin{itemize}
    \item We \textit{employ a divide-and-conquer strategy} to simplify 3D PDN analysis by decomposing it into multiple 2D PDNs, each corresponding to a single die (Fig.~\ref{fig:n-die PDN with unit current}). TSVs are modeled as current sources, enabling independent analysis of each 2D PDN (\cref{fig:tsv-model}).
    \item Subsequently, the proposed framework \textit{computes the intra-die effective resistance between specified points within each 2D PDN}. These intra-die results are essential for capturing the local resistive behavior within each die.
    \item Finally, after completing the intra-die analysis, the framework \textit{calculates the inter-die effective resistances}, specifically from bumps to loads. 
    % as depicted in  Fig.~\ref{fig:tsv-model}. 
    %These inter-die resistances are crucial for evaluating the overall PDN performance.
\end{itemize}   
By employing the above flow, the proposed framework is inherently well-suited for GPU acceleration, enabling rapid evaluation of various design alternatives to ensure that effective resistance is consistently balanced throughout the 3D IC. %A simplified flow chart of the proposed effective resistance computing procedure is illustrated in Fig.~\ref{fig:flow_chart}.

%This approach ultimately helps designers avoid costly revisions and performance issues later in the design process. 

%This approach takes advantage of the regular structure of early-stage PDNs, particularly in terms of their uniformity and the manageable size of the 2D subproblems.

\eat{The working flow of the proposed framework is shown in \cref{fig:flow_chart}. Being fed with the equivalent circuit of 3D PDN, the two inherent algorithms respectively outputs IR drop, TSV currents and effective resistances. \cref{fig:structure} provides a specific example of an equivalent circuit for a two-die 3D PDN, constructed under the following general assumptions:
\begin{itemize}
    \item The 3D PDN is composed of 2D regular resistor networks interconnected with TSVs which are also modeled as resistors. 
    \item Power supplies are connected to the top die which is modeled as ideal independent voltage sources. Loads are connected to the bottom die which is modeled as ideal independent current sources.
\end{itemize}}
\eat{
\begin{figure}[htbp] 
\centerline{\includegraphics[width=0.95\linewidth]{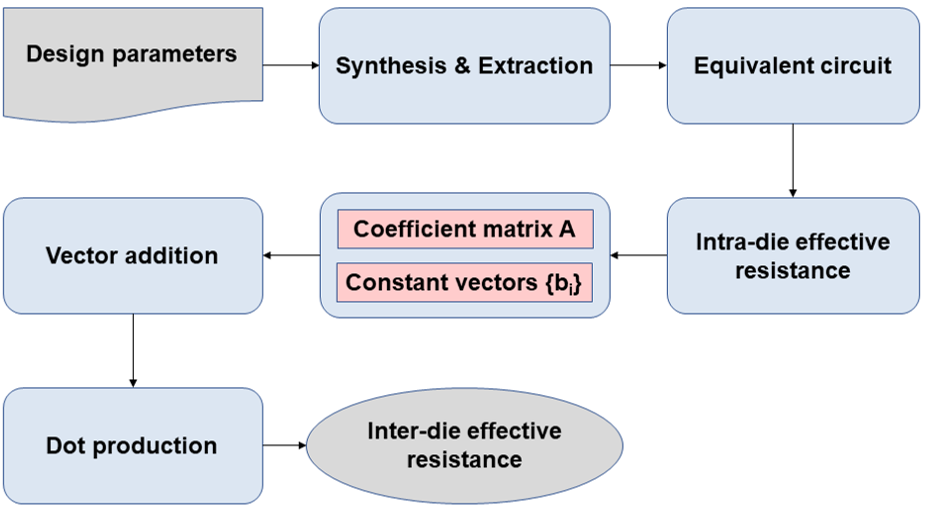}}\vspace{-0.3cm}
\caption{Flow chart of the proposed effective resistance analysis.}
\label{fig:flow_chart}
\end{figure}
}
% 2. the figure of the overall structure
% \begin{figure}[htbp]
% \centerline{\includegraphics[width=0.8\linewidth]{figs/structure.png}}
% \caption{overall structure(rough version).}
% \label{fig:structure}
% \end{figure}

\subsection{Synthesis and Extraction}
In the early stages of 3D IC PDN design, the focus is on efficiently capturing critical layout features, such as topology and bump/TSV planning, rather than replicating the final tape-out design~\cite{early_stage_analysis}. Common assumptions for early-stage 3D IC PDN design include: (1) Bumps connected to the top-tier die are modeled as ideal independent voltage sources, while loads are modeled as ideal independent current sources; (2) The 3D IC PDN is formed by interconnecting multiple 2D PDNs using TSVs. To synthesize the PDN, synthesis algorithms generate the layout based on interconnect specifications predetermined by process technology, metal properties, and design constraints. After synthesis, the initial grid is legalized to ensure compliance with design rules regarding width and pitch. Finally, layout-dependent parasitics, such as resistance, are extracted using standard extraction flows or library-based models~\cite{early_stage_analysis}.

%In the early stages of 3D IC PDN design, the goal is not to replicate the tape-out design but to efficiently capture the critical layout features, such as topology and bump/TSV planning, using regular and homogeneous structures for each module~\cite{early_stage_analysis}. There a few assumptions that are commonly deployed in practical early-stage analysis: (1) Bumps are connected to the top-tier die and modeled as ideal independent voltage sources, while loads are modeled as ideal independent current sources. (2) The 3D IC PDN is interconnected by multiple 2D PDNs using TSVs.

%To achieve a synthesized PDN, fast synthesis algorithms can be employed based on interconnect specifications pre-determined by process technology, metal properties, stack dielectrics, and other design constraints. After synthesis, the initial power grid is legalized to ensure that the width and pitch meet design rules. Finally, layout-dependent parasitics, such as resistance, can be extracted using standard parasitics extraction flows or library-based models~\cite{early_stage_analysis}.

\subsection{Effective Resistance Calculation for A Specific Bump-Load Pair}
\label{Sec:DivideAndConquer}

\begin{table}[t]\vspace{-0.2cm}
\caption{Notations Used in Sec.~\ref{Sec:Framework}.}
    \begin{center}
        \begin{tabular}{|c|c|}
            \hline
            \textbf{Name} & \textbf{Description} \\
            \hline
            $m$ & Number of nodes connecting to voltage sources \\
            \hline
            $n$ & Number of nodes connecting to current loads \\
            \hline
            $d$ & Number of stacked dies \\
            \hline
            $p$ & Number of TSVs \\
            \hline
            $t_j^i$ & Intersection node of $TSV_j$ and the $i_{th}$ tier $die_i$ \\
            \hline
            $s^i$ & (Virtual) supply node of the $i_{th}$ tier $die_i$ \\
            \hline
            $V_j^i$ & Voltage of $t_j^i$ \\
            \hline
            $V_s^i$ & Voltage of $s^i$ \\
            \hline
            $I_k^{i,i+1}$ & Inter-die TSV current between $die_i$ and $die_{i+1}$ for $TSV_k$ \\
            \hline
            $Rt_k^{i,i+1}$ & Inter-die resistance between $die_i$ and $die_{i+1}$ for $TSV_k$ \\
            \hline
            $R_{j,k}^i$ & Intra-die effective resistance between $t_j^i$ and $t_k^i$ \\
            \hline
            $R_{s,k}^i$ & Intra-die effective resistance between $s^i$ and $t_k^i$ \\
            \hline
            %$a_{s,j,k}^i$ & $(R_{s,j}^i + R_{s,k}^i - R_{j,k}^i)/2$ \\
            %\hline
        \end{tabular}
    \end{center}
    \label{tab:notation}\vspace{-0.6cm}
\end{table}
The most accurate method to compute the effective resistance between a bump and a load in a 3D IC is to inject a unit current (\textit{e.g.}, 1A) and measure the resulting voltage difference between the two nodes, as illustrated in \cref{fig:n-die PDN with unit current}. This voltage difference directly represents the effective resistance. However, computing this for a 3D IC involves solving for the voltages of all nodes, making it computationally expensive to evaluate effective resistances for all bump-load combinations.

To facilitate the analysis, a few notations are introduced in Table~\ref{tab:notation}. 
% While the actual bumps are placed on the top tier, we can designate a virtual supply node $s^i$ on the $i_{th}$ tier ${die}_i$ as a reference point in the subsequent derivations. For a given bump-load pair, let the bump and load nodes be $s^1$ and $s^d$, respectively. 
While the actual bumps are placed on the top tier, we can still designate a supply node $s^i$ on the $i_{th}$ tier ${die}_i$. For a given bump-load pair, let bump and load nodes be $s^1$ and $s^d$, respectively. 
The effective resistance can then be expressed as:\vspace{-1ex}
\begin{equation}
    \label{eq:sum_of_vol_diff}
    R = V_s^1 - V_s^d = \sum_{i=1}^{d-1} (V_s^i - V_s^{i+1}).
\vspace{-1ex}\end{equation}
% This indicates that the effective resistance of a bump-load pair in a 3D PDN can be decomposed into \textit{intra-die effective resistances} of multiple 2D PDNs, \textit{i.e.}, $V_s^i - V_s^{i+1}$. 
This indicates that the effective resistance of a bump-load pair in a 3D PDN can be decomposed into several components, each corresponding to the voltage difference between the supply nodes of two adjacent dies, ${i.e.}$, $V_s^i - V_s^{i+1}$.

Based on Sec.~\ref{Sec:Background}, the voltage at any intersection node $t^i_j$ between the $j_{th}$ TSV and the $i_{th}$ die can be derived using Eq.~\eqref{eq:effR based node voltage}\footnote[1]{$I^{0,1}$ and $I^{d,d+1}$ are zero due to the absence of ${die}_0$ and ${die}_{d+1}$.}.
\begin{equation}
    \label{eq:die_i_application}\vspace{-1ex}
    V_j^i = V_s^i - \frac{1}{2}\sum_{k=1}^p (I^{i,i+1}_k - I^{i-1,i}_k)(R_{s,j}^i + R_{s,k}^i - R_{j,k}^i),\vspace{-1ex}
\end{equation}
where $j = 1, 2, \dots, p$ for $p$ TSVs. Since the voltages at the intersection nodes of a TSV (\textit{i.e}., the endpoints of a TSV segment) between two tiers must obey Ohm's law, $V_j^i - V_j^{i+1} = I_j^{i,i+1} R_t^{i,i+1}$, we can subtract the expression for ${die}_{i+1}$ from that for ${die}_i$, for $i = 1, 2, \dots, d-1$:\vspace{-1ex}
\begin{equation}
\label{eq:seqential_subtration}
\begin{aligned}\hspace{-4ex}
I_j^{i,i+1}Rt_j^{i,i+1} & = (V_s^i - V_s^{i+1}) - \sum_{k=1}^{p} I_k^{i,i+1}(a^i_{s,j,k} + a^{i+1}_{s,j,k}) \\
& + \sum_{k=1}^p I_k^{i-1,i} a^i_{s,j,k} + \sum_{k=1}^p I_k^{i+1,i+2} a^{i+1}_{s,j,k},
\end{aligned}\vspace{-1ex}
\end{equation}
where $j = 1, 2, \dots, p$ and $a_{s,j,k}^i=(R_{s,j}^i + R_{s,k}^i - R_{j,k}^i)/2$. 
% In Eq.~\eqref{eq:seqential_subtration}, if the inter-die TSV currents $I_j^{i,i+1}$ are known, the intra-die effective resistance, $i.e.$, $V_s^i - V_s^{i+1}$ can then be easily computed using Eq.~\eqref{eq:effR_for_regular_grid} and \eqref{eq:seqential_subtration}. In addition, as previously assumed, a unit current is injected for effective resistance computation. Thus, the following condition always holds for inter-die TSV currents at different tiers:\vspace{-1ex}
In Eq.~\eqref{eq:seqential_subtration}, only the voltage difference $V_s^i - V_s^{i+1}$ and the inter-die TSV currents $I_j^{i,i+1}$ are unknown, since $a_{s,j,k}^i$ can be easily computed using Eq.~\eqref{eq:effR_for_regular_grid}. In addition, as previously assumed, a unit current is injected for effective resistance computation. Thus, the following condition always holds for inter-die TSV currents at different tiers:\vspace{-1ex}
\begin{equation}
    \label{eq:sum_of_segment_current}
    \sum_{j=1}^p I_j^{i,i+1} = 1,\quad\quad i = 1, 2, \dots, d-1.\vspace{-1ex}
\end{equation}
%where $i = 1, 2, \dots, d-1$. 

With the above derivations, Eq.~\eqref{eq:seqential_subtration} and \eqref{eq:sum_of_segment_current} can be further rewritten in matrix form:
\begin{equation}
\label{eq:reduced_matrix_form}
    \begin{bmatrix}
        A + D & B \\
        B^T & \mathbf{0}
    \end{bmatrix}
    \begin{bmatrix}
        \mathbf{I} \\
        \mathbf{V}
    \end{bmatrix}
    =
    \begin{bmatrix}
        \mathbf{0} \\
        \mathbf{1}
    \end{bmatrix},
\end{equation}
where the sub-matrices $A$, $B$ and $D$ are:
\begin{equation}
    A = 
    \begin{bmatrix}
        \begin{smallmatrix}
        A^1 + A^2 & -A^2 & \mathbf{0} & \hdots & \mathbf{0} \\
        -A^2 & A^2 + A^3 & -A^3 & \hdots & \mathbf{0} \\
        \mathbf{0} & -A^3 & A^3 + A^4 & \hdots & \mathbf{0} \\
        \vdots & \vdots & \vdots & \ddots & \vdots \\
        \mathbf{0} & \mathbf{0} & \mathbf{0} & \hdots & A^{d-1} + A^d
        \end{smallmatrix}
    \end{bmatrix},
\end{equation}

\begin{equation}
    B = 
    \begin{bmatrix}
        \mathbf{1}_p & \hdots & \mathbf{0} \\
        \vdots & \ddots & \vdots \\
        \mathbf{0} & \hdots & \mathbf{1}_p
    \end{bmatrix},
\end{equation}
\begin{equation}
    % D = diag([Rt_1^{1,2}, \dots, Rt_p^{1,2}, Rt_1^{2,3}, \dots, Rt_p^{d-1,d}]),
    D = diag([Rt_1^{1,2}, \dots, Rt_p^{1,2}, \dots, Rt_1^{d-1,d}, \dots, Rt_p^{d-1,d}]),
\end{equation}
%In Eq.~(10), $A^i$ is defined as:
\begin{equation}
    A^i =
    \begin{bmatrix}
        a^i_{s,1,1} & \hdots & a^i_{s,1,p} \\
        \vdots & \ddots & \vdots \\
        a^i_{s,p,1} & \hdots & a^i_{s,p,p}
    \end{bmatrix},
\end{equation}
and $diag(.)$ converts a vector to a diagonal matrix. $\mathbf{1}_p$ is an $p$-dimensional column vector of ones. $\mathbf{I}$ and $\mathbf{V}$ are:\vspace{-1ex}
\begin{equation}
    \mathbf{I} =
    \begin{bmatrix}
        I_1^{1,2} &
        \hdots &
        I_p^{1,2} &
        \hdots &
        I_1^{d-1,d} &
        \hdots &
        I_p^{d-1,d}
    \end{bmatrix}^T,\vspace{-1ex}
\end{equation}
\begin{equation}\label{eq:v_expression}
    \mathbf{V} =
    \begin{bmatrix}
        -(V_s^1 - V_s^2) &
        \hdots &
        -(V_s^{d-1} - V_s^d)
    \end{bmatrix}^T.\vspace{-1ex}
\end{equation}

%In the above expressions:

%Lastly, $D$ is $diag(Rt_1^{1,2}, \dots, Rt_p^{1,2}, Rt_1^{2,3}, \dots, Rt_p^{d-1,d})$.
Thus, the effective resistance for a particular bump/load pair can be rigorously obtained using Eqs.~\eqref{eq:sum_of_vol_diff} and \eqref{eq:reduced_matrix_form}-\eqref{eq:v_expression}. It is noted that \textbf{the dimension of the coefficient matrix in Eq.~\eqref{eq:reduced_matrix_form} is $(p+1) \times (d-1)$}, approximately corresponding to the number of TSVs multiplied by the number of dies, which is \textit{significantly smaller }than the system conductance matrix that must be solved in the traditional method, with dimensions equivalent to the total number of nodes.

\subsection{Fast Computation of Effective Resistances for Arbitrary Bump-Load Combinations}
Eq.~\eqref{eq:reduced_matrix_form} defines the relationship between intra-die and inter-die effective resistance for a specific bump-load pair. However, a major challenge arises: \textbf{changing either the bump or load node requires recomputing the coefficient matrix, $[A + D \quad B; B^T \quad 0]$, in Eq.~\eqref{eq:reduced_matrix_form}}. This significantly impedes the efficient computation of effective resistances across numerous bump-load pairs, as the matrix decomposition cannot be reused. The root cause is the dependency on terms like $R_{s,j}^i$, which vary with the specific bump and load nodes.

\begin{figure}[t]
    \centering
    \vspace{-0.4cm}
    \includegraphics[width=0.8\linewidth]{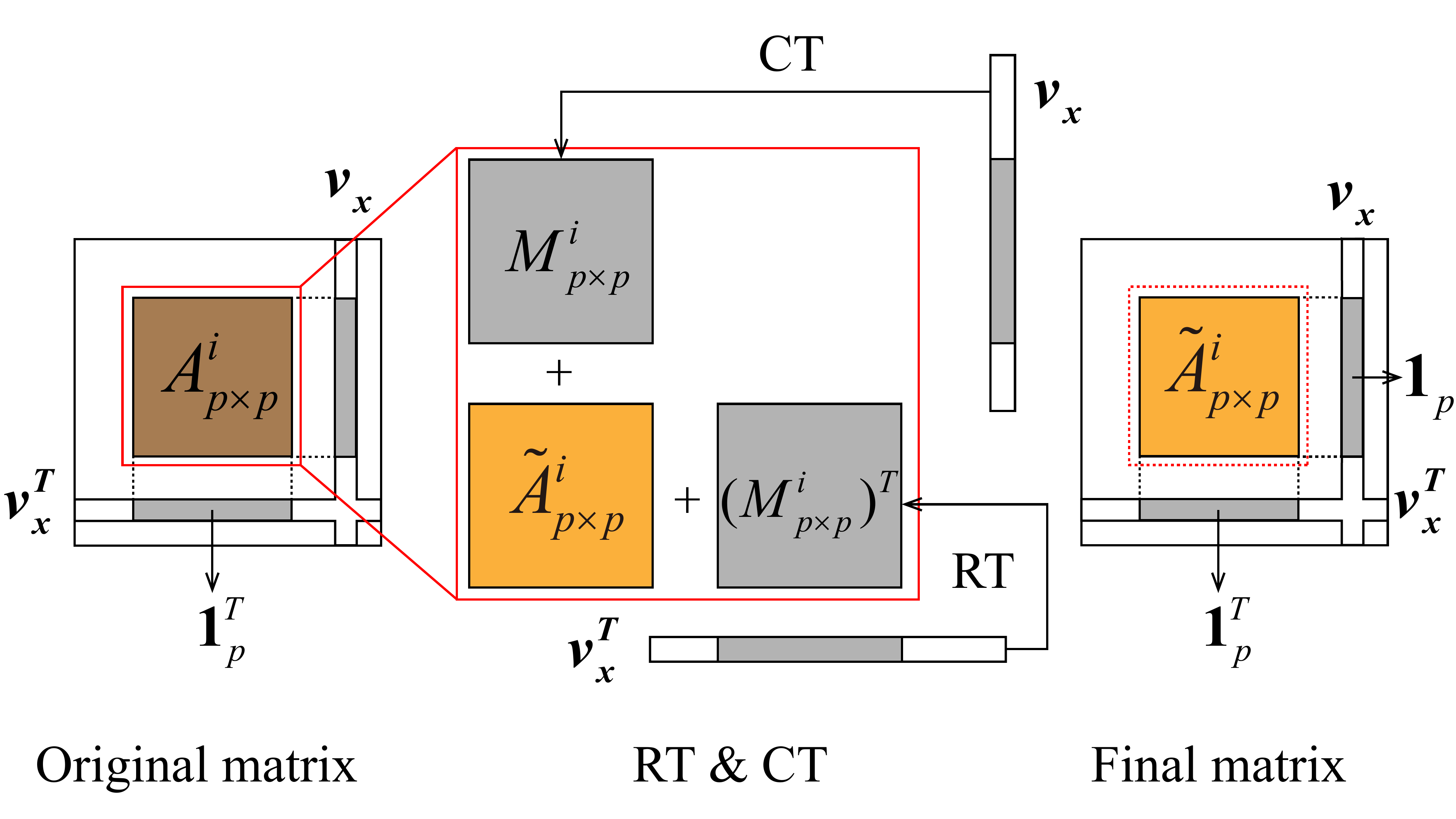}
    % \vspace{-0.5cm}
    \caption{An illustrative example of row transformation.}
    \vspace{-0.5cm}
    \label{fig:example_row_transformation}
\end{figure}

To address this, we apply row and column transformations to the matrix, eliminating dependencies on specific bump-load combinations and producing a reusable coefficient matrix for effective resistance computation. Specifically, the matrix $A^i$ can be expressed as the sum of three components: $\tilde{A}^i + M^i + (M^i)^T$, where 
\begin{equation}
    M^i = 
    \begin{bmatrix}
        R_{s,1}^i\mathbf{1}_p & R_{s,2}^i\mathbf{1}_p & \hdots & R_{s,p}^i\mathbf{1}_p
    \end{bmatrix},
\end{equation}
\begin{equation}
    \tilde{A}^i=-\frac{1}{2}
    \begin{bmatrix}
    0 & R^i_{1,2} & \hdots & R^i_{1,p} \\
    R^i_{2,1} & 0 & \hdots & R^i_{2,p} \\
    \vdots & \vdots & \ddots & \vdots\\
    R^i_{p,1} & R^i_{p,2} & \hdots & 0 \\
    \end{bmatrix}.
\end{equation}
Importantly, only $\tilde{A}^i$ is independent of the bump-load combination. For each $A^i$, we can identify a column $v_x$ and a row $v_x^T$ that align with $\mathbf{1}_p$, as shown in \cref{fig:example_row_transformation}. By performing row transformations (RTs) and column transformations (CTs), we effectively eliminate the contributions of $M^i$ and $(M^i)^T$, resulting in a simplified and reusable coefficient matrix.

%To address this, we apply row and column transformations to the matrix to eliminate these dependencies, thereby obtaining a reusable coefficient matrix for effective resistance computation. Actually, matrix $A^i$ can be expressed as the sum of three matrices: $\tilde{A}^i+M^i+(M^i)^T$, where $M^i=\begin{bmatrix} R_{s,1}^i\mathbf{1}_p & R_{s,1}^i\mathbf{1}_p & \hdots & R_{s,p}^i\mathbf{1}_p\end{bmatrix}$. Notably, only $\tilde{A}^i$ is independent of the bump-load combination. For each $A^i$, we can identify a column $v_x$ and a row $v_x^T$ whose $\mathbf{1}_p$ aligns with it, as illustrated in \cref{fig}. Subsequently, row transformation (RT) and column transformation (CT) can be performed to eliminate $M^i$ and $(M^i)^T$.

% For example, the coefficient matrix $A_e$ for a scenario with two dies and two TSVs is illustrated in \cref{fig:example_row_transformation}, along with the corresponding row and column transformations, where $r_i$ and $c_j$ represent the $i$-th row and $j$-th column of the matrix being transformed. 
Due to the symmetry of the problem, these transformations are applied symmetrically. After the transformations, the structure of the linear system in Eq.~\eqref{eq:reduced_matrix_form} is modified, particularly altering the solution vector. The transformed system can be expressed as:
%For example, the coefficient matrix $A_e$ for a case with two dies and two TSVs is shown in \cref{fig:example_row_transformation}, along with the corresponding row and column transformations where $r_i$ and $c_j$ are the $i$-th row and $j$-th column of the matrix being transformed. Since the problem exhibits symmetry, the transformations are performed symmetrically. After these transformations, the structure of the linear system in Eq.~\eqref{eq:reduced_matrix_form} changes, particularly affecting the solution vector. The transformed system is expressed as:
\begin{equation}
\label{eq:matrix_transformed}
    \begin{bmatrix}
        \tilde{A}+D & B \\
        B^T & \mathbf{0}
    \end{bmatrix}
    \begin{bmatrix}
        \mathbf{I} \\
        \mathbf{\tilde{V}} \\
    \end{bmatrix}
    =
    \mathbf{b},
\end{equation}
where $\mathbf{b}$ is:
\begin{equation}
    \begin{bmatrix}
        \begin{smallmatrix}
                -\frac{1}{2}R_{s,1}^1 & \hdots & -\frac{1}{2}R_{s,p}^1 & \mathbf{0} & -\frac{1}{2}R_{s,1}^{d} & \hdots & -\frac{1}{2}R_{s,p}^d & \mathbf{1}_{p}^T
        \end{smallmatrix}
    \end{bmatrix}^T;
\end{equation}
$\mathbf{\tilde{V}}$ is a vector of $\tilde{v}_i$ with its $i_{th}$ component as:\vspace{-1ex}
\begin{equation}
    \begin{aligned}
    \tilde{v}_i =
            &-(V_s^{i}-V_s^{i+1})+\sum_{j=1}^{d-1}({R_{s,j}^i+R_{s,j}^{i+1}})I_j^{i,i+1}\\
            &-\sum_{j=1}^{d-1}{R_{s,j}^iI_j^{i-1,i}-\sum_{j=1}^{d-1}R_{s,j}^{i+1}}I_j^{i+1,i+2}.
    \end{aligned}\vspace{-1ex}
\end{equation}
Similarly, we can define $R_{s,j}^{0}$ and $R_{s,j}^{d+1}$ as zero, and $\tilde{A}$ is analogous to $A$ but with $A^i$ replaced by $\tilde{A}^i$: 
% \begin{equation}
%     \tilde{A}^i=-\frac{1}{2}
%     \begin{bmatrix}
%     0 & R^i_{1,2} & \hdots & R^i_{1,p} \\
%     R^i_{2,1} & 0 & \hdots & R^i_{2,p} \\
%     \vdots & \vdots & \ddots & \vdots\\
%     R^i_{p,1} & R^i_{p,2} & \hdots & 0 \\
%     \end{bmatrix}.
% \end{equation}

Eq.~\eqref{eq:matrix_transformed} can be written as $C\mathbf{x} = \mathbf{b}$, where the target effective resistance $R_{eff}$ is elegantly expressed as $R_{eff} = -\mathbf{x}^T\mathbf{b}$. Importantly, the coefficient matrix $C$ is independent of the specific bump and load nodes, depending only on geometric and physical parameters of 3D IC. This property \textbf{ensures a unique coefficient matrix, requiring only one decomposition}.

\begin{figure*}[t]
    \centering\vspace{-0.5cm}
    \includegraphics[width=0.9\linewidth]{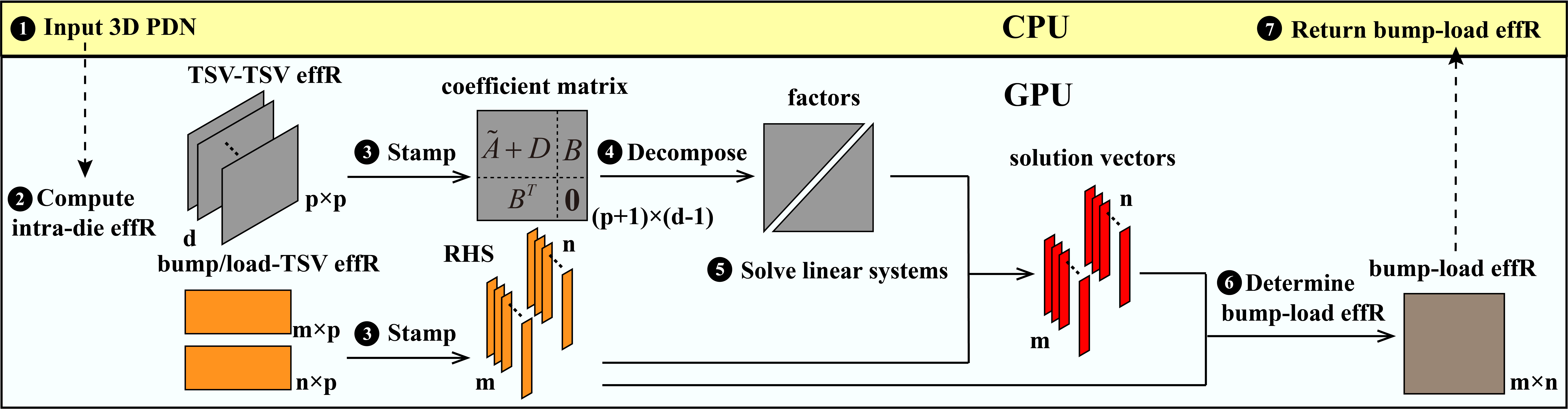}\vspace{-0.3cm}
    \caption{GPU acceleration of the proposed framework for effective resistance (effR) computation.}
    \label{fig:GPU_CPU_overflow}\vspace{-0.3cm}
\end{figure*}
 For $m$ bump nodes and $n$ load nodes, there are $m \times n$ bump-load combinations, which would normally require solving $m \times n$ linear systems based on Eq.~\eqref{eq:matrix_transformed}. However, leveraging the linearity of the system, this number is reduced to $m + n$, as the vector $\mathbf{b}$ can be split into two components:\vspace{-1ex} %Moreover, for $m$ bumps and $n$ loads, there are $m \times n$ combinations, requiring the solution of $m \times n$ linear systems similar to Eq.~\eqref{eq:matrix_transformed}. However, due to the linearity of the system, this number can be reduced to $m + n$, as $\mathbf{b}$ can be expressed as the sum of two vectors:
\begin{equation}
    \label{b_partition}
    \begin{aligned}
        \mathbf{b}=
        &-\frac{1}{2}
        \begin{bmatrix}
            R_{s,1}^1 & \hdots & R_{s,p}^1 & \mathbf{0} & -\mathbf{1}_p^T
        \end{bmatrix}^T \\
        &-\frac{1}{2}
        \begin{bmatrix}
            \mathbf{0} & R_{s,1}^{d} & \hdots & R_{s,p}^d & -\mathbf{1}_p^T
        \end{bmatrix}^T.
    \end{aligned}\vspace{-1ex}
\end{equation}
Let $\mathbf{b} = \mathbf{b}^1 + \mathbf{b}^d$, where $\mathbf{b}^1$ and $\mathbf{b}^d$ depend only on the endpoints in $\text{die}_1$ and $\text{die}_d$, respectively. For $m$ bump nodes in $\text{die}_1$ and $n$ load nodes in $\text{die}_d$, their corresponding vectors are $\mathbf{b}_i^1$ ($i = 1, 2, \dots, m$) and $\mathbf{b}_j^d$ ($j = 1, 2, \dots, n$).
% denote Eq.~\eqref{b_partition} as $\mathbf{b} = \mathbf{b}^1 + \mathbf{b}^d$, where $\mathbf{b}^1$ and $\mathbf{b}^d$ depend only on the endpoints in ${die}_1$ and ${die}_d$, respectively. For $m$ bump nodes in ${die}_1$ and $n$ load nodes in ${die}_d$, we define $\mathbf{b}_i^1$ ($i = 1, 2, \dots, m$) and $\mathbf{b}_j^d$ ($j = 1, 2, \dots, n$).

%We denote \eqref{b_partition} as $\mathbf{b}=\mathbf{b}^1+\mathbf{b}^d$. Apparently, $\mathbf{b}^1$($\mathbf{b}^d$) solely depends on the endpoint of target effective resistance in $die_1$($die_d$). Given $m$ bump nodes in $die_1$ and $n$ load nodes in $die_d$, there are $m$ vectors homogeneous to $\mathbf{b}^1$ and $n$ vectors homogeneous to $\mathbf{b}^d$ which we denoted as $\mathbf{b}_i^1,i\in\{1,2,...,m\}$ and $\mathbf{b}_j^d,j\in\{1,2,...,n\}$ respectively. 

Taking these vectors as constants and $C$ as the coefficient matrix, we can solve $m + n$ linear systems, yielding solution vectors $\mathbf{x}_i^1$ ($i = 1, 2, \dots, m$) and $\mathbf{x}_j^d$ ($j = 1, 2, \dots, n$). The effective resistance between the $i$-th bump in ${die}_1$ and the $j$-th load in ${die}_d$ is then computed as:\vspace{-1ex}
\begin{equation}
    \label{eq:partition of solution}
    R_{eff}(i,j)=-(\mathbf{x}^1_i+\mathbf{x}^d_j)^T(\mathbf{b}^1_i+\mathbf{b}^d_j).\vspace{-1ex}
\end{equation}

%Taking these vectors as constant vectors and $C$ as coefficient matrix, we can obtain $m+n$ linear systems. The solution vectors of these linear systems can also be denoted as $\mathbf{x}_i^1,i\in\{1,2,...,m\}$ and $\mathbf{x}_j^d,j\in\{1,2,...,n\}$. Thereafter, the effective resistance across the $i_{th}$ bump node in $die_1$ and the $j_{th}$ load node in $die_d$ can be computed by 
In summary, to compute $m \times n$ effective resistances between bump and load nodes, we solve $m + n$ linear systems using a small coefficient matrix with a size proportional to the number of dies and TSVs. Constructing these systems requires intra-die effective resistance and TSV segment resistance, both amenable to parallel computation, as shown in Eq.~\eqref{eq:effR_for_regular_grid}. After decomposing the coefficient matrix, all $m + n$ systems can be solved simultaneously, with matrix operations like dot products and additions significantly accelerated. Advanced parallel computing platforms further enhance the framework's efficiency.

\subsection{GPU Acceleration}

\begin{figure} \vspace{-0.3cm}
    \centering
    \includegraphics[width=0.7\linewidth]{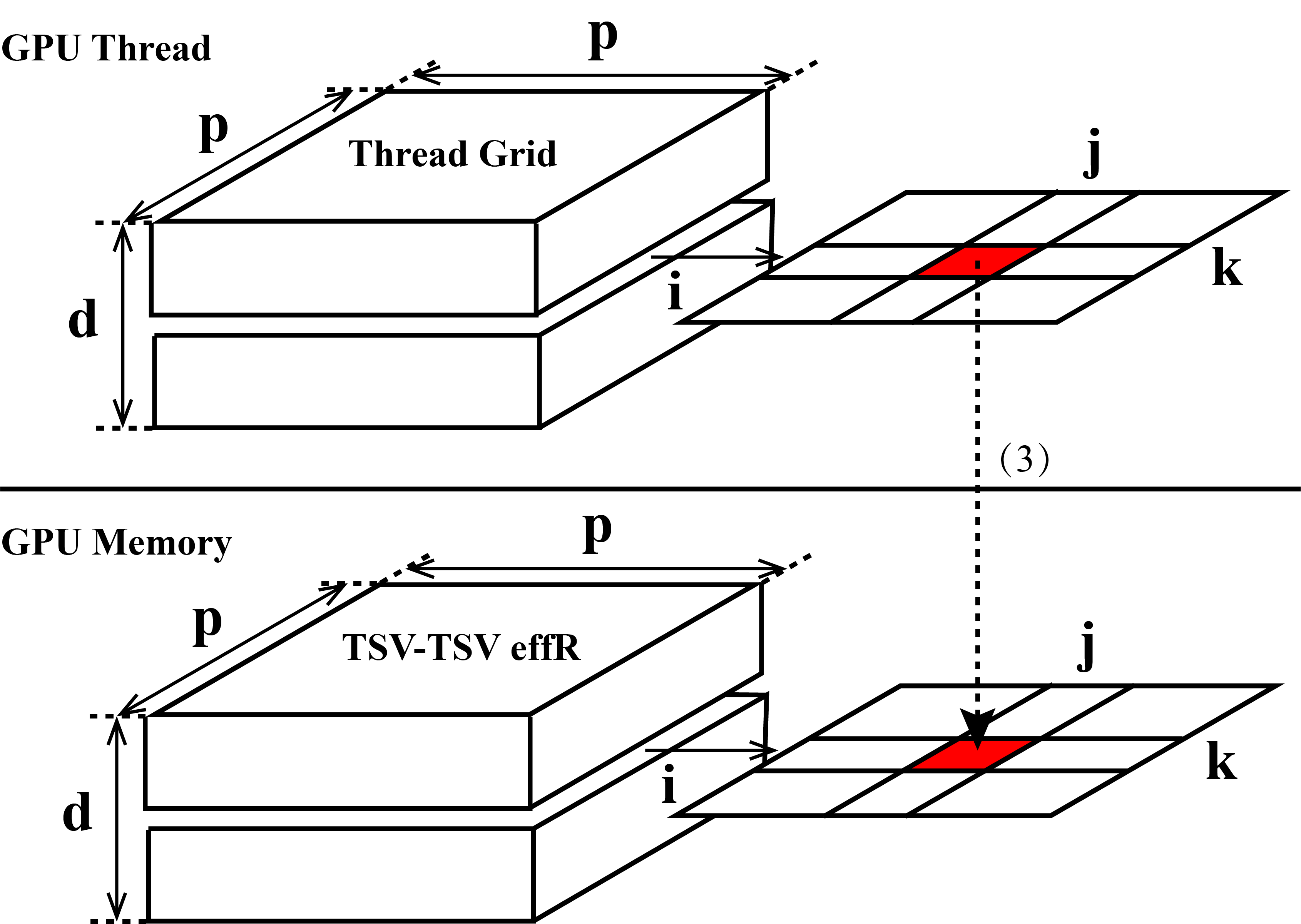}\vspace{-0.3cm}
    \caption{GPU thread/memory management for effective resistance computing.}\vspace{-0.3cm}
    \label{fig:thread_organization}
\end{figure}

While GPUs provide immense parallel computing capabilities with significant potential for effective resistance computation, achieving tangible performance improvements requires a comprehensive GPU-oriented redesign~\cite{gpu_for_pdn}. The proposed framework is specifically engineered to leverage this principle, as summarized in \cref{fig:GPU_CPU_overflow}.
In \cref{fig:GPU_CPU_overflow}, TSV-TSV effective resistances (effR) represent the resistances between TSV nodes and are stored in a 3D array of dimensions $d \times p \times p$. Bump-TSV and Load-TSV effective resistances are stored in two separate arrays of dimensions $m \times p$ and $n \times p$, respectively. These arrays collectively contain all required intra-die effective resistances and are used to construct the unique coefficient matrix and constant vectors. After the coefficient matrix is decomposed, solution vectors are computed using substitution. The bump-load effective resistances are then calculated using vector addition and dot products, utilizing both the constant and solution vectors. The results are organized into an array of dimensions $m \times n$ and transferred back to the CPU host.

As shown in \cref{fig:GPU_CPU_overflow}, the GPU handles the bulk of the computation, while the CPU manages input and output tasks. This division of labor maximizes the GPU's parallel processing capabilities while minimizing data transfer overhead between the CPU and GPU. To further optimize performance, we reformulate Eq.~\eqref{eq:effR_for_regular_grid} and refine the GPU thread configuration to avoid redundant computations and reduce floating-point operations. For thread allocation, taking the computation of TSV-TSV effective resistances as an example (illustrated in \cref{fig:thread_organization}), a straightforward scheme is to allocate $d \times p \times p$ threads. Each thread, indexed by $(i, j, k)$, computes the effective resistance $R_{j,k}^i$ between the $j_{th}$ and $k_{th}$ TSV nodes within ${die}_i$ using Eq.~\eqref{eq:effR_for_regular_grid}. The results are stored in a GPU array of size $d \times p \times p$. However, since $R_{j,k}^i = R_{k,j}^i$, allocating $p \times p$ threads is unnecessary. Instead, only $p \times (p-1)/2$ threads are needed, as the results of threads indexed by $(i, j, k)$ and $(i, k, j)$ are identical, which reduces redundant thread execution.

Additionally, directly computing intra-die effective resistance using Eq.~\eqref{eq:effR_for_regular_grid} introduces significant redundancy. To alleviate the GPU's computational burden, we partition Eq.~\eqref{eq:effR_for_regular_grid} into several smaller components. Each component is computed in parallel on the GPU, and the partial results are summed up afterward. 

Finally, to further optimize performance, a recursive form of the multiple-angle formula is employed for the cosine terms in Eq.~\eqref{eq:effR_for_regular_grid}. This approach minimizes the use of expensive cosine operations on the GPU, replacing them with multiplication operations, which require significantly fewer floating-point computations.
Since many operations in \cref{fig:GPU_CPU_overflow} are matrix-centric, established GPU-accelerated libraries such as CuPy~\cite{nishino2017cupy} can be utilized to improve the implementation efficiency.

\section{Experimental Results}

% \begin{table*}[htbp]
% \centering
% \caption{The runtime of effective resistance analysis for different PDN design}
% \label{tab:runtime}
% \begin{tabular}{|c|c|c|c|c|c|c|c|c|}
% \hline
% PDN & \#nodes & \#dies & \#TSV & \#Bump & \#Load node & Runtime/s & Max abs error & Avg abs error \\ \hline
% 1   & 6.7E6   & 4      & 1.0E3 & 1.0E3  & 1.0E4       & 11.8      &       1.62E-12        &               \\ \hline
% 2   & 6.7E6   & 4      & 3.0E3 & 3.0E3  & 1.0E4       & 14.8      &               &               \\ \hline
% 3   & 1.0E7   & 5      & 4.5E3 & 1.1E3  & 1.0E4       & 16.6      &               &               \\ \hline
% 4   & 1.0E7   & 5      & 4.5E3 & 4.5E3  & 1.0E4       & 18.9      &               &               \\ \hline
% 5   & 1.6E7   & 6      & 6.8E3 & 6.8E3  & 1.0E4       & 33.4      &               &               \\ \hline
% 6   & 3.2E7   & 6      & 6.8E3 & 6.8E3  & 1.0E4       & 43.7      &               &               \\ \hline

% \end{tabular}
% \end{table*}
% Please add the following required packages to your document preamble:
% \usepackage{multirow}
% Please add the following required packages to your document preamble:
% \usepackage{multirow}
\begin{table*}[htbp]
\centering\vspace{-0.6cm}
\caption{Comparison on accuracy and speed for effective resistance calculation using the proposed framework, a golden direct solver~\cite{chen2008algorithm}, and a recent pseudo-inverse-based solver~\cite{effR-sparse-cholesky-inverse} on different early-stage 3D IC PDN benchmarks.}\vspace{-0.2cm}
\label{tab:pdn_benchmark}
\begin{tabular}{|c|c|c|c|c|c|c|ccc|c|c|c|}
\hline
\multirow{2}{*}{Design} & \multirow{2}{*}{\#Die} & \multirow{2}{*}{Footprint ($mm^2$)} & \multirow{2}{*}{\#Nodes} & \multirow{2}{*}{\#TSV} & \multirow{2}{*}{\#Bump} & \multirow{2}{*}{\#Load} & \multicolumn{3}{c|}{Runtime (sec.)}                                                             & \multirow{2}{*}{Speedup}  & \multirow{2}{*}{ARE} & \multirow{2}{*}{MRE} \\ \cline{8-10}
                        &                        &                                    &                          &                        &                         &                         & \multicolumn{1}{c|}{Golden~\cite{chen2008algorithm}} & \multicolumn{1}{c|}{Pseudo~\cite{effR-sparse-cholesky-inverse}}  & \multicolumn{1}{l|}{Ours} &                            &                     &                     \\ \hline
D1                      & 4                      & 5.5$\times$7.3                            & 4.4e6                    & 1.8e4                  & 432                     & 1e4                     & \multicolumn{1}{c|}{3.4e6}         & \multicolumn{1}{c|}{31.0}  & 13.3                      & 2.6e5X\textbackslash{}2.3X & 5.0e-9              & 5.1e-9              \\ \hline
D2                      & 5                      & 5.5$\times$7.3                            & 5.5e6                    & 2.2e4                  & 432                     & 1e4                     & \multicolumn{1}{c|}{4.3e6}         & \multicolumn{1}{c|}{38.7}  & 14.1                      & 3.0e5X\textbackslash{}2.7X & 5.6e-9              & 5.7e-9              \\ \hline
D3                      & 6                      & 5.5$\times$7.3                            & 6.6e6                    & 2.7e4                  & 432                     & 1e4                     & \multicolumn{1}{c|}{5.6e6}         & \multicolumn{1}{c|}{46.6}  & 15.1                      & 3.7e5X\textbackslash{}3.1X & 6.1e-9              & 6.2e-9              \\ \hline
D4                      & 4                      & 7.8$\times$11.9                           & 1.0e7                    & 2.2e4                  & 1040                    & 1e4                     & \multicolumn{1}{c|}{2.2e7}         & \multicolumn{1}{c|}{72.5}  & 17.1                      & 1.3e6X\textbackslash{}4.2X & 8.2e-9              & 8.3e-9              \\ \hline
D5                      & 5                      & 7.8$\times$11.9                           & 1.3e7                    & 2.8e4                  & 1040                    & 1e4                     & \multicolumn{1}{c|}{2.8e7}         & \multicolumn{1}{c|}{95.0}  & 18.9                      & 1.5e6X\textbackslash{}5.0X & 8.2e-9              & 8.2e-9              \\ \hline
D6                      & 6                      & 7.8$\times$11.9                           & 1.5e7                    & 3.4e4                  & 1040                    & 1e4                     & \multicolumn{1}{c|}{3.4e7}         & \multicolumn{1}{c|}{110.3} & 20.8                      & 1.6e6X\textbackslash{}5.3X & 8.8e-9              & 8.9e-9              \\ \hline
D7                      & 4                      & 11.0$\times$11.0                          & 1.3e7                    & 2.6e4                  & 1369                    & 1e4                     & \multicolumn{1}{c|}{4.1e7}         & \multicolumn{1}{c|}{103.2} & 25.8                      & 1.6e6X\textbackslash{}4.0X & 9.4e-08             & 9.5e-08             \\ \hline
D8                      & 5                      & 11.0$\times$11.0                          & 1.7e7                    & 3.3e4                  & 1369                    & 1e4                     & \multicolumn{1}{c|}{4.9e7}         & \multicolumn{1}{c|}{126.3} & 28.5                      & 1.7e6X\textbackslash{}4.4X & 1.7e-08             & 1.8e-08             \\ \hline
D9                      & 6                      & 11.0$\times$11.0                          & 2.0e7                    & 4.0e4                  & 1369                    & 1e4                     & \multicolumn{1}{c|}{NA}            & \multicolumn{1}{c|}{149.6} & 31.5                      & NA\textbackslash{}4.7X     & NA                  & NA                  \\ \hline
\end{tabular}\vspace{-0.5cm}
\end{table*}
 All experiments are conducted on an Ubuntu 22.04 host with an Intel Xeon Platinum 8475B (48 cores, 503 GB RAM) and an NVIDIA A100 GPU (80 GB). In addition to a most recent pseudo-inverse-based solver~\cite{effR-sparse-cholesky-inverse}, a conventional CHOLMOD-based direct solver~\cite{chen2008algorithm} is used as the golden reference.

\subsection{Evaluation of Accuracy and Speed}
\cref{tab:pdn_benchmark} summarizes the parameters of 9 early-stage 3D IC industrial benchmarks based on 14nm technology~\cite{pdn_for_many_tier,3d_pdn_benchmark,HBM}. Columns 2-7 detail the number of dies (\#Die), footprint ($mm^2$), number of PDN nodes (\#Node), TSVs (\#TSV), bumps (\#Bump), and loads (\#Load). For early-stage design, bumps and TSVs are assumed to be uniformly distributed. Columns 8-10 show the runtime for the golden solver~\cite{chen2008algorithm}\footnote[1]{The golden solver cannot complete D9 due to memory limitations.}, the pseudo-inverse-based solver~\cite{effR-sparse-cholesky-inverse}\footnote[2]{The solver~\cite{effR-sparse-cholesky-inverse} trades accuracy for speed using pseudo-inverse methods.}, and the proposed framework. 

Both solvers in \cite{chen2008algorithm,effR-sparse-cholesky-inverse} are serial, with runtimes proportional to the number of bump-load pairs. The pseudo-inverse-based solver~\cite{effR-sparse-cholesky-inverse} achieves speedups at the cost of accuracy, while the proposed framework is unaffected by bump and TSV topology, delivering 5-6 orders of magnitude speedup over the golden solver and 2-5$\times$ speedup over \cite{effR-sparse-cholesky-inverse}, as shown in Column 11. The last two columns list the average and maximum relative errors (denoted as ARE and MRE) compared to the golden solver~\cite{chen2008algorithm}, which are around the orders of $10^{-9}$-$10^{-8}$ and hence negligible. Thus, unlike the prior works~\cite{chen2008algorithm,effR-sparse-cholesky-inverse}, \textbf{the proposed framework achieves both high accuracy and speed}.

\begin{figure}[t]
    \centering
    \subfigure[Runtime v.s. \#Nodes.]{
        \label{fig:time_vs_nodeCnt}
        \includegraphics[width=0.45\linewidth]{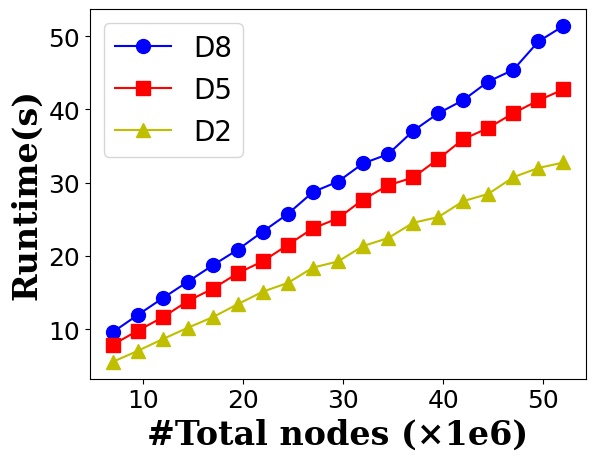}
    }
    \subfigure[Runtime v.s. \#TSVs/Die.]{
        \label{fig:time_vs_TSVCnt}
        \includegraphics[width=0.45\linewidth]{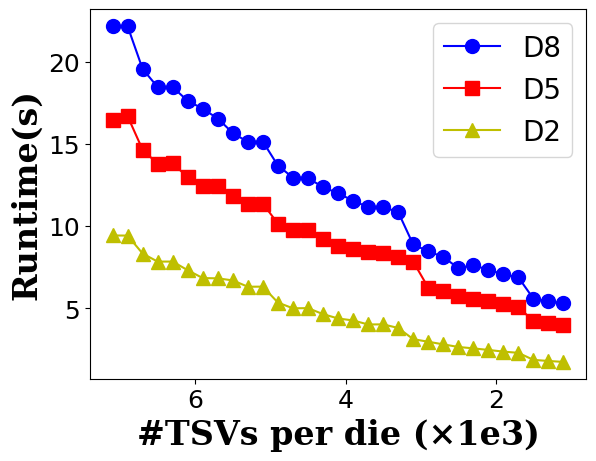}
    }
    \subfigure[Runtime v.s. \#Bump-load pairs.]{
        \label{fig:time_vs_blPair}
        \includegraphics[width=0.45\linewidth]{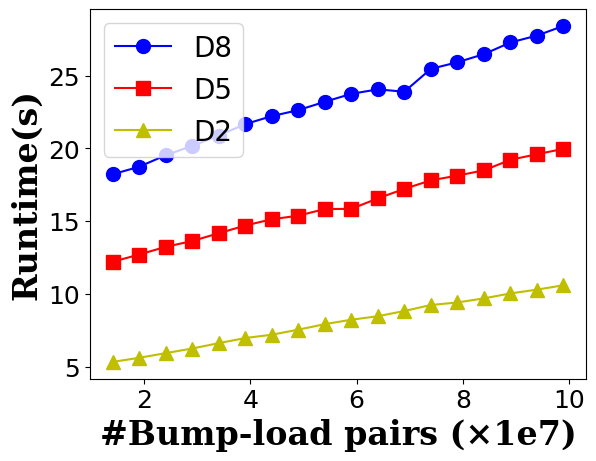}
    }
    \subfigure[Effective resistance distribution.]{
    \label{fig:effR_dist_map}
    \includegraphics[width=0.45\linewidth]{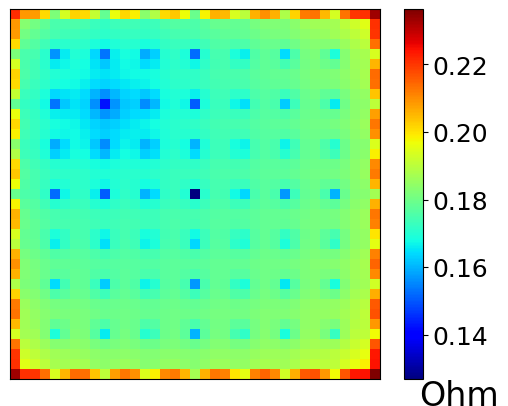}
    }\vspace{-0.2cm}
    \caption{Comparison of (a) runtime v.s. \#Node; (b) runtime v.s. \#TSVs/Die; (c) runtime v.s. \#Bump-load pairs for cases D2, D5 and D7; (d) Contour of effective resistance distribution.}
    \label{fig:runtime_vs_variable}\vspace{-0.3cm}
\end{figure}

%The results show that computing all bump-load effective resistances using CHOLMOD within a reasonable timeframe is nearly infeasible. In contrast, our method enables fast and accurate computation. For the largest case, 'Design 9,' which consists of $2 \times 10^7$ nodes, our framework computes the effective resistances for all $1.3 \times 10^7$ bump-load pairs in just 31.5 seconds.

To evaluate the scalability of our framework, we selected three key parameters, the total number of nodes, the number of TSVs per die, and the number of bump-load pairs, and then varied each parameter using `D2,' `D5,' and `D8' as baselines while keeping all other parameters constant. The relationships between runtime and these parameters are summarized in \cref{fig:time_vs_nodeCnt,fig:time_vs_TSVCnt,fig:time_vs_blPair}. It is worth noting that while the runtime increases linearly with those key parameters, \textbf{the growth rate of the runtime is significantly smaller compared to the growth rate of these parameters}. This again demonstrates the efficiency and scalability of the proposed framework in handling large-scale 3D PDN benchmarks.

\begin{figure}[t]
    \centering\vspace{-0.0cm}
    % \hspace{-0.4cm}
    % \subfigure[Band-like pattern]{
    %     \label{fig:HBM-like pattern}
    %     \includegraphics[width=0.29\linewidth]{figs/test_circuit.png}
    % }
    % \subfigure[uniform pattern]{
    %     \label{fig:uniform pattern}
    %     \includegraphics[width=0.4\linewidth]{figs/test_circuit_uniformTSV.png}
    % }
    % \subfigure[bump current distribution map]{
    %     \label{fig:bcur dist map}
    %     \includegraphics[width=0.8\linewidth]{figs/current_map_combined.png}
    % }
    \subfigure[Band-like placement and its current distribution]{
        \includegraphics[width=0.8\linewidth]{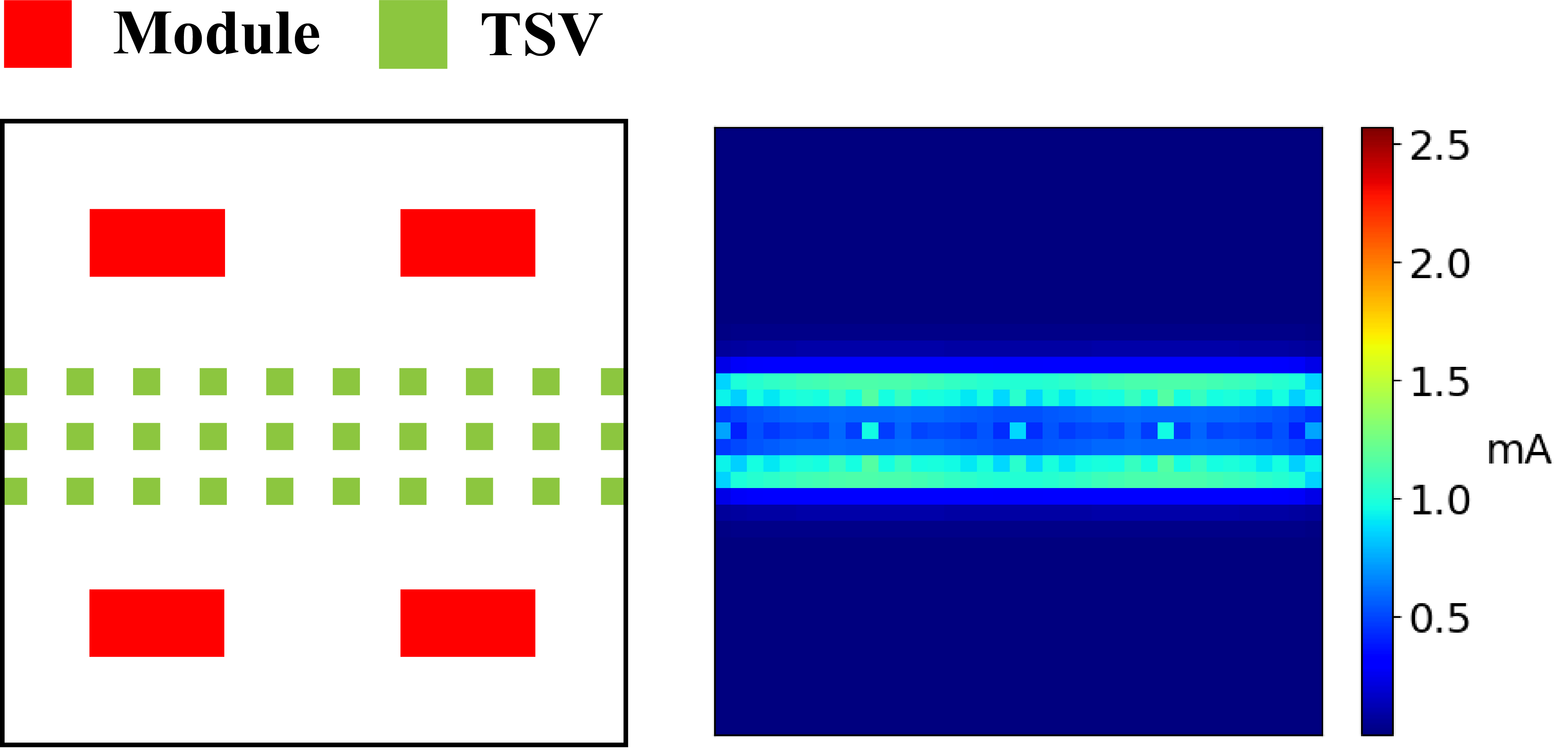}
    }
    \subfigure[Uniformly-distributed placement and its current distribution]{
        \includegraphics[width=0.8\linewidth]{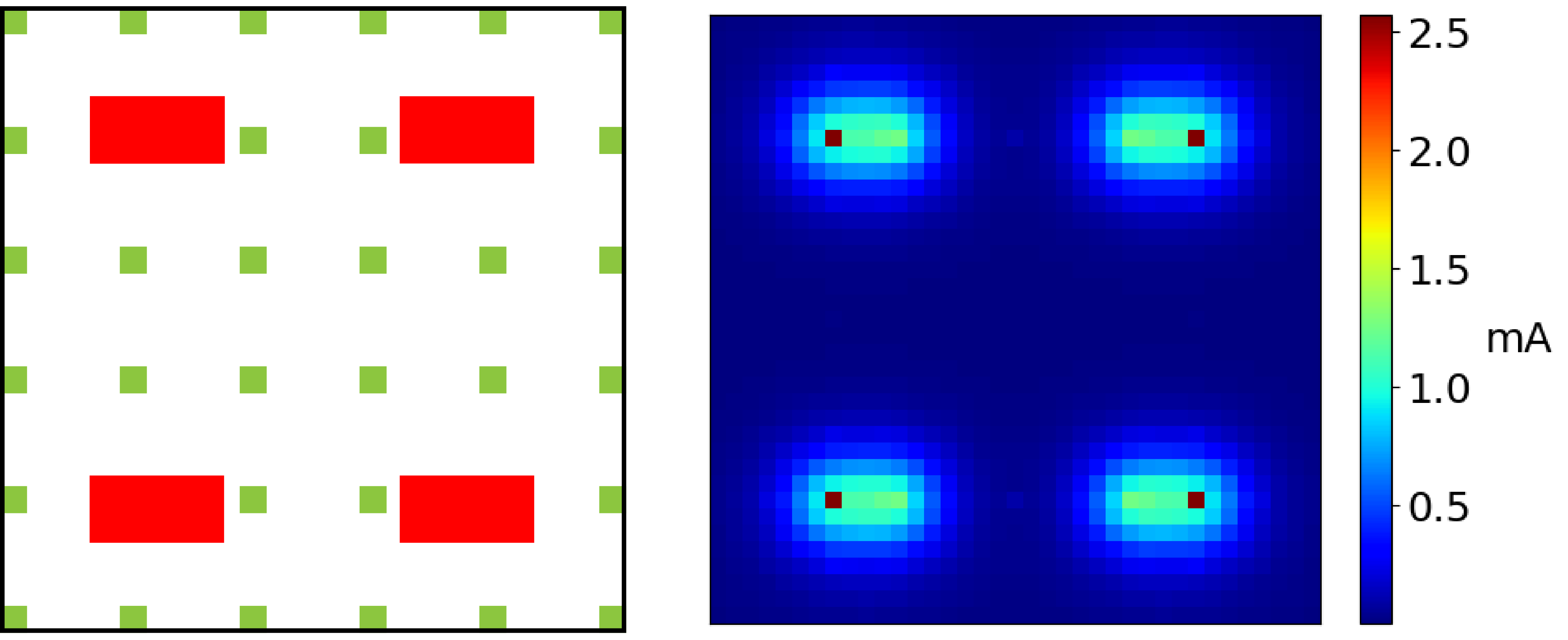}
    }
    \vspace{-0.3cm}
    \caption{Through-bump current distributions for different TSV planning options: (a) Band-like placement; (b) Uniformly-distributed placement.}
    \label{fig:pattern_vs_cur_dist}\vspace{-0.3cm}
\end{figure}

\subsection{TSV Planning Using the Proposed Framework}
The proposed framework can be readily used to facilitate evaluations of different design options at the early design stage. For example, designers can examine the contour of the effective resistance distribution in \cref{fig:effR_dist_map} to select the most suitable TSV planning option. \cref{fig:pattern_vs_cur_dist}(a) and (b) illustrate the impact of different TSV planning strategies, both of which are based on practical design guidelines. \cref{fig:pattern_vs_cur_dist}(a) (left) places all TSVs in a centrally clustered configuration to enable a band-like connection, a setup commonly used in mainstream HBM products, whereas \cref{fig:pattern_vs_cur_dist}(b) (left) distributes TSVs uniformly across the entire die to facilitate better current and thermal distribution. The through-bump currents are demonstrated and compared for the two scenarios. For the same four functional blocks, each occupying $2.35\times1.18 mm^2$ with a current density of $220 A/cm^2$, the band-like connection in \cref{fig:pattern_vs_cur_dist}(a) results in 23.4\% of the total bumps carrying approximately 95\% of the total current, while also contributing to significant routing area savings. In contrast, the uniformly distributed connection  in \cref{fig:pattern_vs_cur_dist}(b) results in 53.2\% of the total bumps carrying 95\% of the total current. These insights allow designers to make informed trade-offs between different application scenarios to select the optimal TSV planning strategy.

\section{Conclusion}
\label{sec:conclusion}
The proposed framework presents a highly efficient and scalable solution for effective resistance analysis in the early design stages of 3D IC PDNs. By leveraging the parallel processing capabilities of GPUs, the method achieves substantial speedup compared to traditional approaches, making it feasible to handle large-scale 3D PDN designs with millions to billions of nodes. The results show that our approach not only provides fast and accurate resistance computation but also facilitates efficient design options exploration in 3D IC development.

\bibliographystyle{IEEEtran}
\bibliography{3DPowerPlanner}

@INPROCEEDINGS{effR_approach_for_PG,
  author={Köse, Selçuk and Friedman, Eby G.},
  booktitle={Proceedings of 2010 IEEE International Symposium on Circuits and Systems}, 
  title={Fast algorithms for power grid analysis based on effective resistance}, 
  year={2010},
  volume={},
  number={},
  pages={3661-3664},
  doi={10.1109/ISCAS.2010.5537772}}

@article{Wu_2004, title={Theory of resistor networks: the two-point resistance}, volume={37}, ISSN={0305-4470, 1361-6447}, DOI={10.1088/0305-4470/37/26/004}, abstractNote={The resistance between two arbitrary nodes in a resistor network is obtained in terms of the eigenvalues and eigenfunctions of the Laplacian matrix associated with the network. Explicit formulae for two-point resistances are deduced for regular lattices in one, two and three dimensions under various boundary conditions including that of a Mo¨bius strip and a Klein bottle. The emphasis is on lattices of ﬁnite sizes. We also deduce summation and product identities which can be used to analyse large-size expansions in two and higher dimensions.}, number={26}, journal={Journal of Physics A: Mathematical and General}, author={Wu, F Y}, year={2004}, month=jul, pages={6653–6673}, language={en} }

@INPROCEEDINGS{Kozhaya_Nassif_Najm_2001,
  author={Kozhaya, J.N. and Nassif, S.R. and Najm, F.N.},
  booktitle={2001 IEEE/ACM International Conference on Computer-Aided Design (ICCAD)}, 
  title={Multigrid-like technique for power grid analysis}, 
  year={2001},
  volume={},
  number={},
  pages={480-487},
  doi={10.1109/ICCAD.2001.968685}}

@INPROCEEDINGS{3D_Intel,
  author={Black, Bryan and Annavaram, Murali and Brekelbaum, Ned and DeVale, John and Jiang, Lei and Loh, Gabriel H. and McCaule, Don and Morrow, Pat and Nelson, Donald W. and Pantuso, Daniel and Reed, Paul and Rupley, Jeff and Shankar, Sadasivan and Shen, John and Webb, Clair},
  booktitle={2006 39th Annual IEEE/ACM International Symposium on Microarchitecture (MICRO'06)}, 
  title={Die Stacking (3D) Microarchitecture}, 
  year={2006},
  volume={},
  number={},
  pages={469-479},
  doi={10.1109/MICRO.2006.18}}

@INPROCEEDINGS{random_walk,
  author={Haifeng Qian and Nassif, S.R. and Sapatnekar, S.S.},
  booktitle={Proceedings 2003. Design Automation Conference (IEEE Cat. No.03CH37451)}, 
  title={Random walks in a supply network}, 
  year={2003},
  volume={},
  number={},
  pages={93-98},
  doi={10.1109/DAC.2003.1218831}}

@INPROCEEDINGS{early_stage_analysis,
  author={Zhuo, Cheng and Gan, Houle and Wei-Kai Shih},
  booktitle={2014 51st ACM/EDAC/IEEE Design Automation Conference (DAC)}, 
  title={Early-stage power grid design: Extraction, modeling and optimization}, 
  year={2014},
  volume={},
  number={},
  pages={1-6},
  doi={10.1145/2593069.2593129}}

@ARTICLE{hierarchical_analysis,
  author={Zhao, M. and Panda, R.V. and Sapatnekar, S.S. and Blaauw, D.},
  journal={IEEE Transactions on Computer-Aided Design of Integrated Circuits and Systems}, 
  title={Hierarchical analysis of power distribution networks}, 
  year={2002},
  volume={21},
  number={2},
  pages={159-168},
  doi={10.1109/43.980256}}

@ARTICLE{GPU_based_PDN_analysis,
  author={Feng, Zhuo and Zhao, Xueqian and Zeng, Zhiyu},
  journal={IEEE Transactions on Computer-Aided Design of Integrated Circuits and Systems}, 
  title={Robust Parallel Preconditioned Power Grid Simulation on GPU With Adaptive Runtime Performance Modeling and Optimization}, 
  year={2011},
  volume={30},
  number={4},
  pages={562-573},
  doi={10.1109/TCAD.2010.2091437}}

@ARTICLE{effR_of_2layers,
  author={Kose, Selçuk and Friedman, Eby G.},
  journal={IEEE Transactions on Circuits and Systems II: Express Briefs}, 
  title={Effective Resistance of a Two Layer Mesh}, 
  year={2011},
  volume={58},
  number={11},
  pages={739-743},
  doi={10.1109/TCSII.2011.2168016}}

@INPROCEEDINGS{Design_of_reliable_3D_IC,
  author={Hung, Shao-Chun and Chakrabarty, Krishnendu},
  booktitle={2020 Design, Automation \& Test in Europe Conference \& Exhibition (DATE)}, 
  title={Design of a Reliable Power Delivery Network for Monolithic 3D ICs}, 
  year={2020},
  volume={},
  number={},
  pages={1746-1751},
  doi={10.23919/DATE48585.2020.9116570}}

@INPROCEEDINGS{2D-3D-PDN-comparison,
  author={Khan, Nauman H. and Alam, Syed M. and Hassoun, Soha},
  booktitle={2009 IEEE International Conference on 3D System Integration}, 
  title={System-level comparison of power delivery design for 2D and 3D ICs}, 
  year={2009},
  volume={},
  number={},
  pages={1-7},
  doi={10.1109/3DIC.2009.5306539}}

@INPROCEEDINGS{Analysis_of_IR_drop,
  author={Ajami, A.H. and Banerjee, K. and Mehrotra, A. and Pedram, M.},
  booktitle={Fourth International Symposium on Quality Electronic Design, 2003. Proceedings.}, 
  title={Analysis of IR-drop scaling with implications for deep submicron P/G network designs}, 
  year={2003},
  volume={},
  number={},
  pages={35-40},
  doi={10.1109/ISQED.2003.1194706}}

@ARTICLE{TSV-Based-3-D-ICs,
  author={Lu, Tiantao and Serafy, Caleb and Yang, Zhiyuan and Samal, Sandeep Kumar and Lim, Sung Kyu and Srivastava, Ankur},
  journal={IEEE Transactions on Computer-Aided Design of Integrated Circuits and Systems}, 
  title={TSV-Based 3-D ICs: Design Methods and Tools}, 
  year={2017},
  volume={36},
  number={10},
  pages={1593-1619},
  doi={10.1109/TCAD.2017.2666604}}

@INPROCEEDINGS{effR-sparse-cholesky-inverse,
  author={Liu, Zhiqiang and Yu, Wenjian},
  booktitle={2023 Design, Automation \& Test in Europe Conference \& Exhibition (DATE)}, 
  title={Computing Effective Resistances on Large Graphs Based on Approximate Inverse of Cholesky Factor}, 
  year={2023},
  volume={},
  number={},
  pages={1-6},
  doi={10.23919/DATE56975.2023.10137201}}

@INPROCEEDINGS{XGBIR,
  author={Pao, Chi-Hsien and Su, An-Yu and Lee, Yu-Min},
  booktitle={2020 Design, Automation \& Test in Europe Conference \& Exhibition (DATE)}, 
  title={XGBIR: An XGBoost-based IR Drop Predictor for Power Delivery Network}, 
  year={2020},
  volume={},
  number={},
  pages={1307-1310},
  doi={10.23919/DATE48585.2020.9116327}}

@INPROCEEDINGS{IncPIRD,
  author={Ho, Chia-Tung and Kahng, Andrew B.},
  booktitle={2019 IEEE/ACM International Conference on Computer-Aided Design (ICCAD)}, 
  title={IncPIRD: Fast Learning-Based Prediction of Incremental IR Drop}, 
  year={2019},
  volume={},
  number={},
  pages={1-8},
  doi={10.1109/ICCAD45719.2019.8942110}}

@ARTICLE{effR-approx-infinity-mirror-technique,
  author={Bairamkulov, Rassul and Friedman, Eby G.},
  journal={IEEE Transactions on Circuits and Systems I: Regular Papers}, 
  title={Effective Resistance of Finite Two-Dimensional Grids Based on Infinity Mirror Technique}, 
  year={2020},
  volume={67},
  number={9},
  pages={3224-3233},
  doi={10.1109/TCSI.2020.2985652}}

@INPROCEEDINGS{effR-for-PG-TSI,
  author={Liu, En-Xiao and Cubillo, Joseph Romen and Li, Er-Ping and Zhao, Huapeng and Oo, Zaw Zaw and Lee, Hui Min},
  booktitle={2012 IEEE Electrical Design of Advanced Packaging and Systems Symposium (EDAPS)}, 
  title={Effective resistance approach for DC analysis of power grid on through-silicon interposer (TSI)}, 
  year={2012},
  volume={},
  number={},
  pages={1-4},
  doi={10.1109/EDAPS.2012.6469410}}

@article{effR_definition,
 ISSN = {00361445},
 URL = {http://www.jstor.org/stable/20454062},
 author = {Arpita Ghosh and Stephen Boyd and Amin Saberi},
 journal = {SIAM Review},
 number = {1},
 pages = {37--66},
 publisher = {Society for Industrial and Applied Mathematics},
 title = {Minimizing Effective Resistance of a Graph},
 urldate = {2024-09-12},
 volume = {50},
 year = {2008}
}

@article{nishino2017cupy,
  title={Cupy: A numpy-compatible library for nvidia gpu calculations},
  author={Nishino, ROYUD and Loomis, Shohei Hido Crissman},
  journal={31st confernce on neural information processing systems},
  volume={151},
  number={7},
  year={2017}
}

@inproceedings{3d_pdn_benchmark, address={Stateline Nevada USA}, title={Benchmarking for research in power delivery networks of three-dimensional integrated circuits}, ISBN={978-1-4503-1954-6},  booktitle={Proceedings of the 2013 ACM International symposium on Physical Design}, publisher={ACM}, author={Luo, Pei-Wen and Zhang, Chun and Chang, Yung-Tai and Cheng, Liang-Chia and Lee, Hung-Hsie and Sheu, Bih-Lan and Su, Yu-Shih and Kwai, Ding-Ming and Shi, Yiyu}, year={2013}, month=mar, pages={17–24}, language={en} }

@INPROCEEDINGS{pdn_for_many_tier,
  author={Healy, Michael B. and Lim, Sung Kyu},
  booktitle={2010 Proceedings 60th Electronic Components and Technology Conference (ECTC)}, 
  title={Power delivery system architecture for many-tier 3D systems}, 
  year={2010},
  volume={},
  number={},
  pages={1682-1688},
  doi={10.1109/ECTC.2010.5490753}}

@INPROCEEDINGS{HBM,
  author={Park, Joonsang and Kim, Seongguk and Son, Keeyoung and Kim, Haeyeon and Kim, Hyunwoo and Kim, Hyunsik and Choi, Seonguk and Kim, Jihun and Kim, Joungho},
  booktitle={2023 IEEE 32nd Conference on Electrical Performance of Electronic Packaging and Systems (EPEPS)}, 
  title={Design and Analysis of an Irregular-Shaped Power Distribution Network (PDN) for High Bandwidth Memory (HBM) Interposer}, 
  year={2023},
  volume={},
  number={},
  pages={1-3},
  doi={10.1109/EPEPS58208.2023.10314861}}

@ARTICLE{gpu_for_pdn,
  author={Feng, Zhuo and Zeng, Zhiyu and Li, Peng},
  journal={IEEE Transactions on Very Large Scale Integration (VLSI) Systems}, 
  title={Parallel On-Chip Power Distribution Network Analysis on Multi-Core-Multi-GPU Platforms}, 
  year={2011},
  volume={19},
  number={10},
  pages={1823-1836},
  doi={10.1109/TVLSI.2010.2059718}}

@article{chen2008algorithm,
  title={Algorithm 887: CHOLMOD, supernodal sparse Cholesky factorization and update/downdate},
  author={Chen, Yanqing and Davis, Timothy A and Hager, William W and Rajamanickam, Sivasankaran},
  journal={ACM Transactions on Mathematical Software (TOMS)},
  volume={35},
  number={3},
  pages={1--14},
  year={2008},
  publisher={ACM New York, NY, USA}
}

@INPROCEEDINGS{direct_solver_for_pg,
  author={He, Qing and Au, William and Korobkov, Alexander and Venkateswaran, Subramanian},
  booktitle={2014 IEEE International Symposium on Electromagnetic Compatibility (EMC)}, 
  title={Parallel power grid analysis using distributed direct linear solver}, 
  year={2014},
  volume={},
  number={},
  pages={866-871},
  doi={10.1109/ISEMC.2014.6899089}}

\end{document}